\iftrue  

\documentclass[conference,a4paper]{APSIPA2021}
\usepackage{amsmath}
\usepackage{graphicx}
\usepackage{multirow}
\usepackage{threeparttable}
\usepackage[backend=biber,style=ieee,]{biblatex}
\usepackage{geometry}
\usepackage{fancyhdr}

\fancypagestyle{firststyle}{
  \fancyhf{}
   \fancyhead[C]{2023 Asia Pacific Signal and Information Processing Association Annual Summit and Conference (APSIPA ASC)}
}

\usepackage{subcaption}  
\usepackage{color}

\else
\documentclass[12pt,a4j,dvipdfmx]{article} 
\usepackage[dvipdfmx]{graphicx}  
\usepackage{amssymb}
\usepackage{amsmath}
\usepackage{float}
\usepackage{color}
\usepackage{comment}
\usepackage[labelformat=parens]{subcaption}
\usepackage{tipa} 
\renewcommand{\baselinestretch}{1.5}

\fi

\begin{document}

\title{Auditory Representation Effective \\ 
 for Estimating Vocal Tract Information}

\author{%
\authorblockN{%
Toshio Irino\authorrefmark{1} and
Shintaro Doan\authorrefmark{1}
}
\authorblockA{%
\authorrefmark{1}
Faculty of Systems Engineering, Wakayama University, Japan \\
E-mail: irino@wakayama-u.ac.jp, doan.shintaro@g.wakayama-u.jp }
%
%
}

\maketitle
\thispagestyle{firststyle}  
\pagestyle{fancy}


\begin{abstract}
We can estimate the size of the speakers based on their speech sounds alone. We had proposed an auditory computational theory of the Stabilised Wavelet-Mellin Transform (SWMT), which segregates information about the size and shape of the vocal tract and glottal vibration, to explain this observation. It has been shown that the auditory representation or excitation pattern (EP) associated with a weighting function based on the SWMT, termed the ``SSI weight,'' can account for the psychometric functions of size perception. In this study, we investigated whether EP with SSI weight can accurately estimate vocal tract lengths (VTLs) which were measured by magnetic resonance imaging (MRI) in male and female subjects. 
It was found that the use of SSI weight significantly improved the VTL estimation. Furthermore, the estimation errors in the EP with the SSI weight were significantly smaller than those in the commonly used spectra derived from the Fourier transform, Mel filterbank, and WORLD vocoder. It was also shown that the SSI weight can be easily introduced into these spectra to improve the performance.

\end{abstract}


\section{Introduction}

\color{black}

We can recognize phonemes pronounced by children, women, and men despite the large differences in their heights. This indicates that our auditory system can extract and identify phonemes in which variations in the pattern of formant frequencies distinguish vowel types and the fundamental frequency determines the pitch. Speaker information can also be extracted simultaneously. 
Speech sounds contain information about vocal tract size, which is closely correlated with speaker size \cite{fitch1999morphology}.
Many psychoacoustic studies have been conducted on size discrimination and phoneme perception from voiced and unvoiced speech sounds (e.g., \,\cite{smith2005processing,ives2005discrimination,smith2007discrimination}; see review in \cite{matsui2022modelling}). 

Irino and Patterson \cite{irino2002segregating} proposed the Stabilised Wavelet-Mellin Transform (SWMT) as a computational theory to explain how the auditory system estimates size and shape of vocal tract separately from glottal pulse information. 
As explained in Section\,\ref{sec:AuditoryProcess}, size estimation from voiced sounds is more difficult than from unvoiced sounds.  This is because voiced speech sounds contain information about vocal tract response (filter characteristics) and glottal vibration (source characteristics) as shown in the source filter theory of speech\,\cite{fant1970acoustic,fant1981source}. Therefore, it is necessary to effectively separate this information.
An auditory model based on SWMT was proposed to explain the experimental results on size discrimination of both unvoiced and voiced speech sounds\cite{irino2017auditory, matsui2022modelling}.  It was demonstrated that introduction of a simple weighting function, referred to as ``size-shape image weight (SSI weight)'' (see Section \,\ref{sec:AuditoryProcess}), enabled to explain the results successfully.  
However, the domain of previous studies was restricted to the explanation of psychometric functions derived using synthetic speech sounds. 
For practical applications in signal processing, it is necessary to demonstrate that the auditory model can also extract vocal tract information from natural speech sounds.

In the current study, we focus on the estimation of vocal tract lengths (VTLs) of different speakers. 
This is because the VTLs of sustained vowels can be accurately measured from magnetic resonance imaging (MRI) data\,\cite{masaki1999mri,takemoto2004method} and thus provide ``ground truth'' usable for evaluation. 
VTL has sometimes been estimated by ``auditory motivated'' models, including  Mel-Frequency Cepstrum Coefficient (MFCC)\,\cite{davis1980comparison,lee1996speaker,pitz2005vocal,sanand2007linear,sarkar2009text}. Recent DNN studies\,\cite{jaitly2013vocal,serizel2014vocal, tan2021vocal} have also used MFCC. Although the computational power has allowed for high VTL estimation accuracy, little question has been raised as to whether MFCC is really effective for vocal tract information estimation, including speech feature extraction. We consider an effective auditory representation to answer this question.


In this paper, we first describe the problem of VTL estimation using the auditory spectrum and our approach. Then, we describe the VTL estimation method and present the evaluation results.
We compared the auditory spectrum with the commonly used Fourier spectrum, the Mel filterbank spectrum\,\cite{davis1980comparison}, and the WORLD vocoder spectrum\,\cite{morise2016world}. 
We also investigated whether introducing the SSI weight into these spectra improves the estimation accuracy.
If it is the case, this simple function, which does not require any training data as in DNN, could improve the performance of many speech processing tasks.


\section{Auditory process to estimate size}
\label{sec:AuditoryProcess}


\subsection{Physics of vocal tract and estimation}
\label{sec:PhysicsSpeech}
The main difference between the vowels of males and females lies in the differences in the VTL and fundamental frequency $F_o$. 
When VTL is shortened by a factor of $1/\alpha$, the formant frequencies $F_1$ and $F_2$ move upward to $\alpha F_1$ and $\alpha F_2$. On the logarithmic frequency axis, $\log F_1$ and $\log F_2$ move up to \{$\log F_1 + \log \alpha$\} and \{$\log F_2 + \log \alpha$\}, respectively. This implies that the logarithmic scale factor, $\log \alpha$, is a constant independent of the formant frequencies.
Therefore, the VTL ratio can be  estimated using the cross-correlation of the log spectra corresponding to the original and shorted vocal tracts. 
Auditory spectra derived by  gammatone or gammachirp filterbanks\,\cite{patterson1995time,irino2006dynamic} are suitable for this purpose because the frequency axis, $\rm ERB_{N number}$, is approximately a log frequency axis above 500\,Hz\,\cite{moore2013introduction}. 
%
This appears to be an easy task if the spectrum is calculated solely from the impulse response of the vocal tract.
However, voiced sounds used in speech communication are derived 
from the convolution of the impulse response of the vocal tract and the waveform of glottal vibration\,\cite{fant1970acoustic}. 
This makes the estimation more difficult than expected.

\begin{figure}[t] 
  \begin{center} 
    \includegraphics[width=0.9\columnwidth]{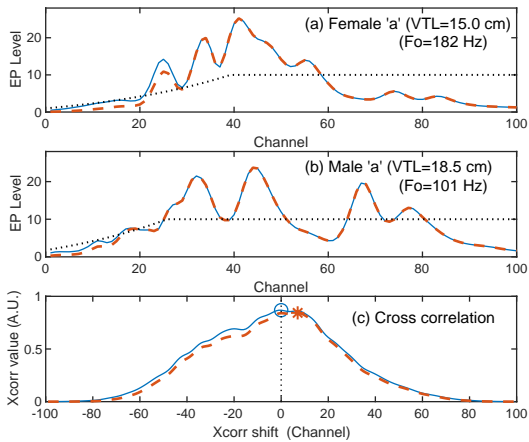}
    \vspace{-8pt}
   \caption {
   Excitation patterns (EPs, i.e., GCFB outputs) of the vowel `a' of a male (a) and a female (b) and their cross-correlation function (c). The horizontal axis for (a) and (b) is the number of GCFB channels. 
 The vertical axis represents the EP level. In (c), the horizontal axis represents the amount of shift in terms of the channel and
the vertical axis represents the correlation function value (arbitrary unit). Blue solid line: EP; black dotted line: SSI weight (see Section \ref{sec:F0adaptiveWeightingFunc}); red dashed line: SSI-weighted EP. Circle (o) and asterisk (*) in (c) indicate the peaks in the cross correlations of the EPs and SSIweighted EPs, respectively.
   }
	 \label{fig:SynthVowel_ExcitationPattern}
  \end{center} 
   \vspace{-10pt}
\end{figure}

\subsection{Problem when using auditory spectrum}
\label{sec:ProblemVTLestimation}

To be more specific, an actual female voice `a' (VTL = 15.0\,cm,  $F_o$ = 182\,Hz) and a male voice `a' (VTL = 18.5\,cm,  $F_o$ = 101\,Hz), drawn from a
database described in Section \ref{sec:MeasuredVTL}, were analyzed
with a dynamic compressive gammachirp auditory filterbank (GCFB)\,\cite{irino2006dynamic,irino2023hearing, GitHub_amlab}.
The output level was averaged over a short period. This is commonly called an excitation pattern (EP)\,\cite{moore2013introduction}.
The solid blue lines in Fig.~\ref{fig:SynthVowel_ExcitationPattern} show the EPs for the female voice (a) and male voice (b). The horizontal axis is the GCFB channel number, which is equally spaced from 100 to 8000 Hz on the ${\rm ERB_N number}$ axis\,\cite{moore2013introduction}; this axis is effectively a logarithmic frequency axis above 500\,Hz.
In the female voice shown in Fig. ~\ref{fig:SynthVowel_ExcitationPattern}\,(a), prominent peaks are observed in channels 24, 32,
and 41. Among these, the peak at channel 41 is associated with a formant, and the spectral shape at higher
frequencies is important for VTL estimation. In contrast, the peaks at channels 24 and 32 are associated with
the harmonics of $F_o$ (182\,Hz), which are resolved harmonics\,\cite{moore2013introduction}. These additional peaks
reduce the accuracy of the VTL estimation. In contrast, for the male voice in Fig. ~\ref{fig:SynthVowel_ExcitationPattern}(b), the peaks corresponding to the resolved harmonics
are relatively small, and four formant peaks are clearly observed. The cross-correlation between the two
EPs is shown by the blue solid line in Fig. ~\ref{fig:SynthVowel_ExcitationPattern}(c).
The peak (blue circle) is obtained at a shift of zero, 
which implies that the VTLs are the same. 
This estimation is obviously incorrect because
the measured VTL ratio is 1.23 (=18.5/15.0). The problem lies in the fact that the resolved harmonics interfere with  estimation. A similar and worse problem occured in the Mel spectrum.



\begin{figure}[t] 
\begin{center}
\scalebox{1.0}{\includegraphics[width=0.8\columnwidth]{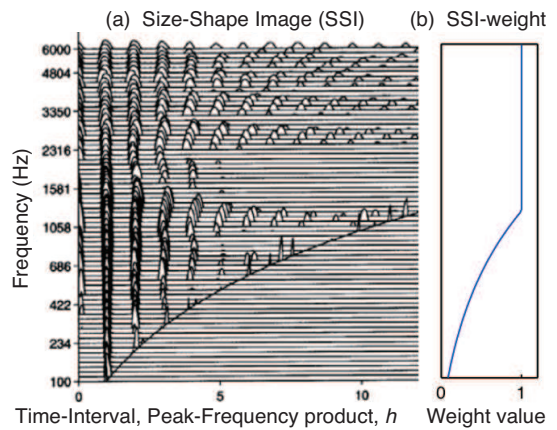}}
\end{center}
\vspace{-10pt} 
\caption{
Size-Shape Image and SSI weight (adapted from \cite{irino2017auditory}). (a) SSI of the vowel `o' \cite{irino2002segregating}. The horizontal axis represents the product of the time interval and the peak frequency of the auditory filter. The vertical axis is the peak frequency of the auditory filter equally spaced on the  $\rm ERB_N number$ axis. (b) Weighting function based on active region of SSI.
(see also Fig.\,\ref{fig:SWMT_SWT_SSIweight} of Appendix A). 
}
\label{fig:SSIweightMaking}
\vspace{-15pt}
\end{figure}

\subsection{Approach from auditory computational theory}
\label{sec:ApproachAuditory}

We approached this problem using a computational theory in which the auditory system can segregate and extract
information about the vocal tract and glottal vibration from speech sounds. Specifically, Irino and Patterson \,\cite{irino2002segregating} proposed the ``Stabilised Wavelet-Mellin Transform'' (SWMT) which has been supported by several psychological experiments on size perception (e.g., \cite{smith2005processing,irino2012comparison,irino2017auditory, matsui2022modelling}).  
%
The SWMT process is briefly described here (see \,\cite{irino2002segregating} for details and the upper path in Fig.\,\ref{fig:SWMT_SWT_SSIweight} of Appendix A). In SWMT, the EP derived from GCFB is converted into a two-dimensional ``Auditory Image (AI)'' by ``Strobe
Temporal Integration (STI),'' which is synchronized with the glottal pulse. The representation of vocal tract response is repeated at the glottal pulse rate (i.e., $F_o$). One cycle of the response is extracted from the AI to obtain a two-dimensional ``Auditory Figure (AF)'', which maximally represents information about a single pulse response of the vocal tract. This representation does not exactly correspond to the impulse response but is much closer than the usual spectrum representation obtained from repeated pulse excitation.
Then, the AF is transformed into a ``size-shape image (SSI),'' as shown in Fig. ~\ref{fig:SSIweightMaking}(a).  Here, the vertical axis represents the peak frequency of the auditory filter, $f_p$, and the horizontal axis is $h$, which is the product of the time interval and peak frequency. SSI is a representation of the single-pulse response and eliminates the response of the adjacent pulse, which is located below the diagonal upright curve.
Although it is a good method in principle, developing an STI algorithm to obtain stable images for various speech sounds is not easy. This is because estimating the strobe point or period synchrony information in the EP is a difficult task due to the variability of speech sounds (e.g., \,\cite{irino2006speech,hohmann2021period}).

\subsection{$F_o$ adaptive weighting function, SSI weight}
\label{sec:F0adaptiveWeightingFunc}
To solve this stability problem, we designed a weight function that corresponds to the area of the active part of the SSI (see also the middle block and the lower path in Fig.\,\ref{fig:SWMT_SWT_SSIweight} of Appendix A). 
Figure\,\ref{fig:SSIweightMaking} (b) shows a function named ``SSI weight'' \,\cite{irino2017auditory, matsui2022modelling}.
This one-dimensional function is directly applicable to EP via simple multiplication. 
The SSI weight ($w_{SSI}$) is defined as
\setlength{\abovedisplayskip}{2pt} 
\setlength{\belowdisplayskip}{2pt} 
\begin{eqnarray}
	w_{SSI}(f_p,F_o) &=& \min(\frac{f_p}{h_{max} \cdot F_o},1),
	\label{eq:SSIweight}
\end{eqnarray}
where $h_{max}$ is the upper limit of $h$ on the horizontal axis of the SSI and determines the area of information extracted from the SSI, as shown in Fig. ~\ref{fig:SSIweightMaking}(a). $f_p$ is the peak frequency of the analysis filer. $F_o$ is the fundamental frequency at the time of the analysis. When $F_o$ is not determined, as in unvoiced sounds, setting $F_o \simeq 0$ will result in a $w_{SSI}$ of unity across the peak frequency. 
The SSI weight was used to explain the results of the human size perception experiments\,\cite{irino2017auditory,matsui2022modelling} and was introduced to a new objective speech intelligibility measure, GESI, to predict the simulated hearing loss sounds\,\cite{irino2022speech,yamamoto2023intelligibility}.
Most importantly, the SSI weight is applicable to any spectral representation because it is a simple weighting function on the frequency axis.



\subsection{Applying the SSI weight to EP}
\label{sec:ApplySSIweight}

In Fig.\,\ref{fig:SynthVowel_ExcitationPattern}, we apply the SSI weight (black dotted line)  to the original spectrum (blue solid line) to derive the weighted spectrum (red dashed line). In particular, in Fig. ~\ref{fig:SynthVowel_ExcitationPattern}(a), the first peak of the resolved harmonics of the female voice is effectively suppressed. 
The red dashed line in Fig.~\ref{fig:SynthVowel_ExcitationPattern}(c) shows the cross-correlation function between the SSI-weighted EPs. The peak (red asterisk) is at a shift of 6, which is approximately 1.2 times the frequency and coincides with the VTL ratio (=1.23) between the male and female. Thus, introducing the SSI weight is expected to reduce the difficulty of VTL estimation.

\section{VTL estimation method and evaluation}
\label{sec:VTLestimation}

\subsection{Measured VTL data}
\label{sec:MeasuredVTL}
The effectiveness of the SSI weight was evaluated using the ``ATR vowel speech MRI data'' \cite{masaki1999mri,takemoto2004method}, which contains five vowel sounds, $v\,\{v| `a',`i',`u',`e',`o'\}$, accompanied by accurate VTLs measured from MRI images of sustained vocalizations. As it contained only 13 male prepared data points, we additionally derived six female data points\,\cite{kitamura2005individual,kitamura2008acoustic}. Therefore, we used data from 19 speakers.

\subsection{Spectral analysis by GCFB}
\label{sec:GCFBanalysis}

We used the GCFB\,\cite{irino2006dynamic,irino2023hearing, GitHub_amlab} for VTL estimation, as described in Section \ref{sec:ApproachAuditory}. It was set to process 100
channels with a sampling frequency of 48 kHz and a filter center frequency range of 100 -- 8000\,Hz. Upon audio input, the EP is output with a frame period of 0.5 ms, 
and the EP spectrogram is calculated. EPs in the range of $\pm$25\,ms from the center of the speech data were averaged to obtain the spectral representation ``$Ep$'' of the vowel described above. The spectral
representation ``$Ep_{SSI}$'' was also calculated by applying the SSI weight in Eq. \ref{eq:SSIweight} with the $F_o$ 
estimated by WORLD\,\cite{morise2016world}.



\subsection{VTL estimation algorithm}
\label{sec:VTLestimationAlgorithm}
For each of the five vowels, the cross-correlation functions of the EPs were calculated for all combinations of the 19 speakers ($N=19$). Between the $i$-th and $j$-th speakers, the shift in the peak position from the center, $c_{ij}$, was extracted using the cross-correlation function.  Note that the EP was linearly interpolated in advance by a factor of 10 to make the resolution of the peak shift as a 0.1 channel.
The resulting shift was assumed to be caused by the VTL difference.
When the orders of $i$ and $j$ are swapped, the amount of shift is the same in absolute value and only the sign is reversed. Therefore, the matrix notation of all permutations is represented as 

\vspace{6pt}
$  
\begin{bmatrix}
0       & c_{12}   & c_{13}   & \cdots & c_{1N}\\
-c_{12} & 0        & c_{23}   & \cdots & c_{2N}\\
-c_{13} & -c_{23}  & 0        & \cdots & c_{3N}\\
\vdots  &  \vdots  &  \vdots  & \ddots & \vdots\\
-c_{1N} &  -c_{2N} &  -c_{3N} & \cdots & 0
\end{bmatrix}.
$
\vspace{6pt}

The relative shift for the $i$-th speaker, $S_i$, from the average of all speakers is calculated by taking the difference between the vertical and horizontal sums of this matrix.
\setlength{\abovedisplayskip}{2pt} 
\setlength{\belowdisplayskip}{2pt} 
\begin{eqnarray}
C_j^{row}  =  \sum_{i=1}^{N}c_{ij},  ~~~~
C_i^{col}  =  \sum_{j=1}^{N}c_{ij} 
\end{eqnarray}
\begin{eqnarray}
S_i & = & (C_i^{row} - C_i^{col})/2N 
\end{eqnarray}
Although this equation is simple, the
equivalent value was obtained using an estimation method with a generalized inverse matrix \cite{irino2013vocal,kawahara2014vocal}. 
Since this value is a shift quantity on a logarithmic scale, its exponent yields the VTL, $L_i$ as
\setlength{\abovedisplayskip}{2pt} 
\setlength{\belowdisplayskip}{2pt} 
\begin{eqnarray}
L_i & = & \exp(q S_i) \cdot \bar L
\end{eqnarray}
where $\bar L$ is the measured VTL averaged across speakers. $q$ is a conversion coefficient that depends on the spectral representation. 
In this study, $q$ was determined to minimize the squared error between the regression line, calculated from the measured and estimated VTLs for all 19 speakers and five vowels, and the 1:1 identical line.

\subsection{Estimation results}

\begin{figure}[t]
\vspace{-10pt}
\hspace{-10pt}
	\begin{minipage}[h]{0.5\columnwidth}
		\begin{center}
                \includegraphics[clip, width=1.1\columnwidth]{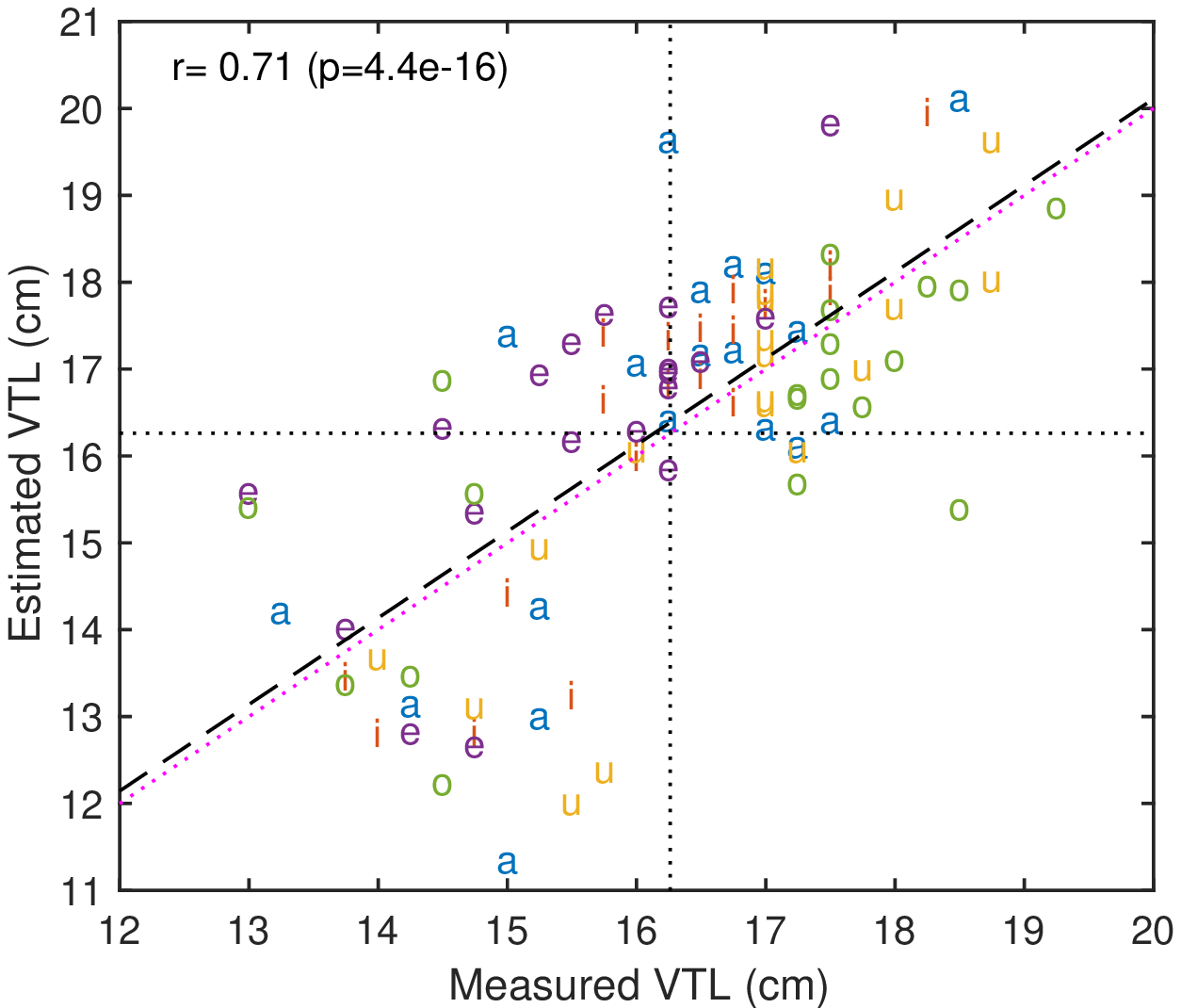}
		\end{center}
	\end{minipage}%
	\begin{minipage}[h]{0.5\columnwidth}
		\begin{center}
			\includegraphics[clip, width=1.1\columnwidth]       {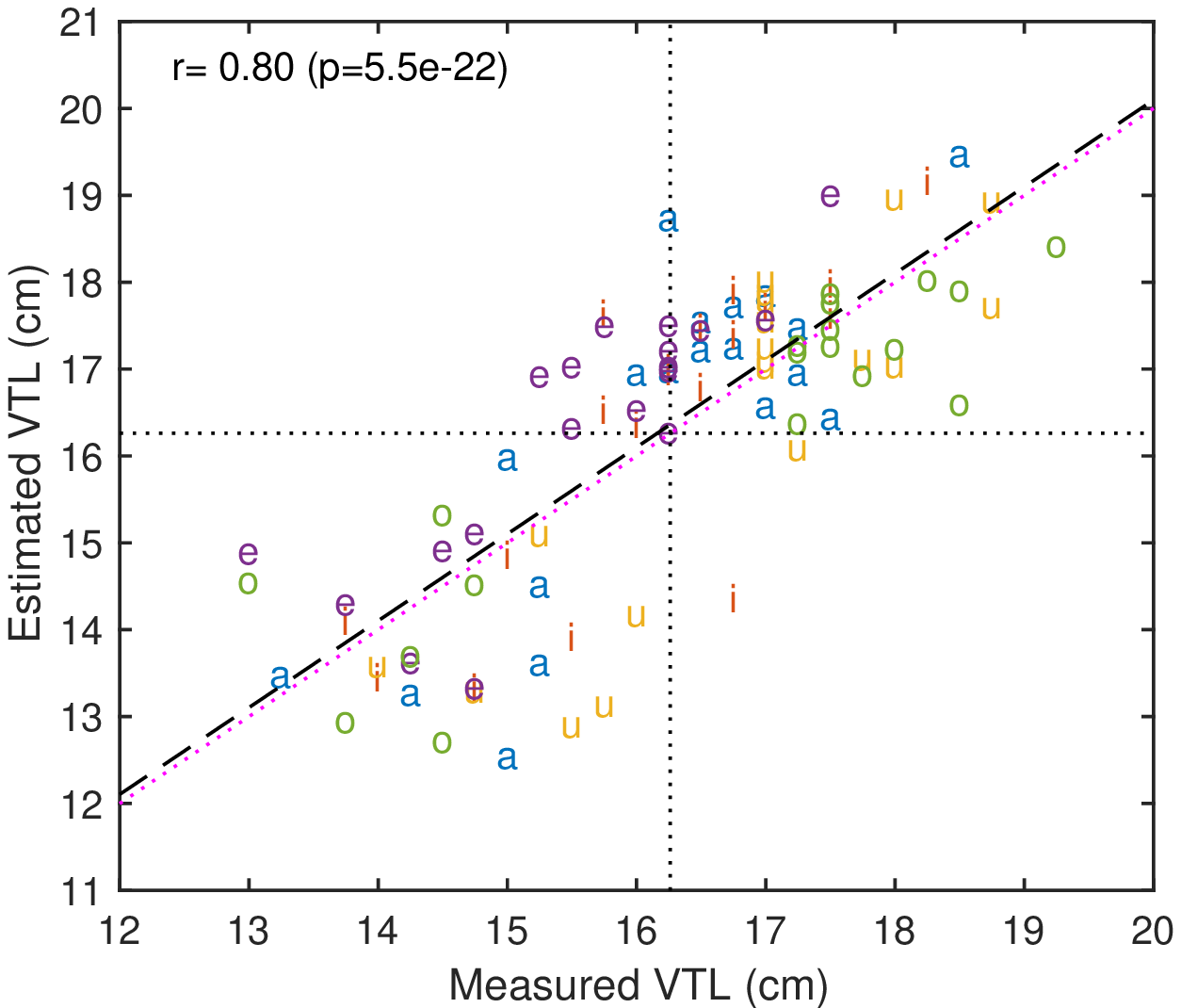}
		\end{center}
	\end{minipage}
	\caption{ Scatter plot of estimated VTL ratio against VTL measured from MRI data. The regression line was obtained using all the vowel and speaker data (dashed lines). Left: $Ep$, right: $Ep_{SSI}$ with $h_{max} = 3.5$.
    Color coded for each vowel. Dotted line: 1:1 identical line. 
  }   \label{fig:VTLscatterRegress} 
    
\end{figure}

Figure ~\ref{fig:VTLscatterRegress} shows scatter plots between the measured  and  estimated VTLs when
using $Ep$ (Fig.\,\ref{fig:VTLscatterRegress}a) and $Ep_{SSI}$ with $h_{max}=3.5$ (Fig.\,\ref{fig:VTLscatterRegress}b). The correlation coefficients for $Ep$ and $Ep_{SSI}$ were calculated for all vowels and speakers. The values were 0.71 and 0.80, respectively.
Vowel labels are generally closer to the regression line in $Ep_{SSI}$ than in $Ep$.
This finding indicates that applying the SSI weight to the EP can improve the accuracy of the VTL estimation, as described in Section \ref{sec:ApplySSIweight}.

\begin{figure}[t]
  \vspace{-10pt}
  \begin{center}
        \includegraphics[width=0.9\columnwidth]{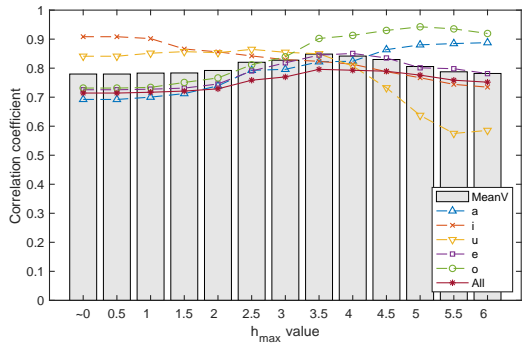}
   \end{center}
    \vspace{-5pt}
  \caption{ Correlation coefficients between measured and estimated VTLs for various $h_{max}$ values. 
  Lines: Coefficients for the individual vowel 
 (`a',`i',`u',`e',`o') and when using all vowels (``All''). Bar: mean value for the five vowels.
  }
        \label{fig:EffectSSIhmax}
         \vspace{-10pt}
\end{figure}

\subsubsection{The effect of $h_{max}$ on VTL estimnation}
Figure \ref{fig:EffectSSIhmax} shows the correlation coefficients between the measured and estimated VTLs calculated for $h_{max}$ values between 0 and 6 in 0.5 steps. We assumed that the frequency response of the vocal tract filter may be best extracted by properly setting the $h_{max}$ value to reduce the resolved harmonics in the EP shown in Fig. \ref{fig:EffectSSIhmax}, although
the $h_{max}$ value was arbitrarily fixed at 5.0 when explaining the results of the human size perception experiments \cite{irino2017auditory,matsui2022modelling}.
It is also important to estimate vocal tract information equally well for all vowel types.

The correlation coefficients were calculated for each vowel,
as shown by the lines in Fig.\,\ref{fig:EffectSSIhmax}.
When $h_{max}$ was less than 3, the correlation coefficients for `i' and `u' (low first formant, $F_1$) were high, while those for `a,' `e,' and `o' (high $F_1$) were low. This result implies that the $F_o$ information from the resolved harmonics was not sufficiently suppressed in the low-frequency region.
In contrast, when $h_{max}$ exceeded 4, the correlation coefficients for `a' and
`o' (high $F_1$) were high, and those for `i' and `u' (low $F_1$) were low. Therefore, $F_1$ and $F_o$ information were excessively suppressed. 
The best results were obtained when $h_{max}$ was 3.5, for which the difference between the five vowels was small, and the correlation coefficient obtained from all five vowels (`All') was the highest. In this case, information about the vocal tract and glottal vibration seemed to be properly separated regardless of vowel type.

\section{Comparison with commonly used spectra}
We compared the estimation performances when using the above auditory representation and commonly used spectral representations to evaluate their effectiveness.
A comparison was made with a Fourier spectrum, ``$F$,'' and a Mel-filterbank (MFB) spectrum\,\cite{davis1980comparison},``$M$.'' We also included a WORLD spectrum\,\cite{morise2016world},``$W$,'' because it reduces the effect of $F_o$ by smoothing the frequency distribution and is commonly used in voice conversion as a successor of Tandem-STRAIGHT\cite{kawahara2008tandem}.
The estimation algorithm was identical to that described in Section \ref{sec:VTLestimationAlgorithm}.

\subsection{Calculation of the spectrum}
\label{sec:CompetitiveSpecRep}
The short-time Fourier spectrogram of the vowel sound was calculated with a frame length of 25\,ms, a hamming window, and a frame shift of 5 ms. The Mel spectrogram was obtained from the STFT spectrogram with a Mel filterbank with 25 filters equally spaced on a Mel frequency axis corresponding to between 100 and 8000\,Hz. The WORLD spectrogram was obtained with a default frame rate of 5\,ms.
Then the obtained amplitude spectrogram, $S(f_p,\tau)$, was subjected to logarithmic
compression, $20\,\log_{10}\{S(f_p,\tau)\}$, and power compression, $S(f_p,\tau)^{P}~~\{P \,|\, 0.1\le P \le 1.0,\, {\rm every}\, 0.1 \}$.
The compressed spectrogram in the range of $\pm$25\,ms from the center was averaged to obtain the spectrum as calculated in EP (Section \ref{sec:GCFBanalysis}).
For the Fourier and WORLD spectra, the frequency axis was logarithmically transformed, and
a spectrum with 100 channels equally spaced between $\log_{10}(100)$ and $\log_{10}(8000)$ was obtained by linear interpolation.
The Mel spectrum with 100 channels was obtained from the 25-channel spectrum via linear interpolation. 

As decribed in section\,\ref{sec:F0adaptiveWeightingFunc}, the SSI weight improved the estimation performance when applied to the EP and was applicable to any spectrum. Therefore, the  SSI weight was also applied to these spectra to investigate its effect. Table \ref{table:SpecAbbreviation} presents the abbreviations for each compressed spectrum.

\begin{table}[t]
  \caption{Abbreviations of the compressed spectrum}
  \label{table:SpecAbbreviation}
  \centering
\begin{tabular}{lcccc}
    \hline
    spectrum  & log & power, $P$ & log+SSI & $P$+SSI\\
    \hline \hline
   Fourier  & $F^{log}$  & $F^{P}$ & $F_{SSI}^{log}$ & $F_{SSI}^{P}$\\
   Mel  & $M^{log}$  & $M^{P}$ & $M_{SSI}^{log}$ & $M_{SSI}^{P}$\\
   WORLD  & $W^{log}$  & $W^{P}$ & $W_{SSI}^{log}$& $W_{SSI}^{P}$\\
    \hline
  \end{tabular}
\end{table}

\begin{figure}[t]
      \begin{subfigure}[b]{\columnwidth}
         \centering
         \hspace{-10pt}
         \includegraphics[width=1\columnwidth]{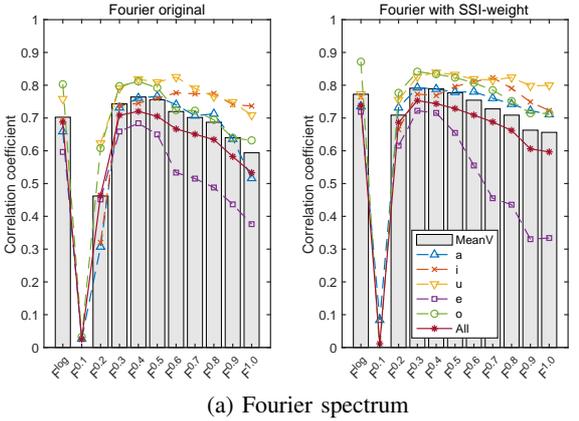}
         \vspace{-5pt}
         \caption{Fourier spectrum}
         \label{fig:CC-CompressFourier}
     \end{subfigure}
         \begin{subfigure}[b]{\columnwidth}
         \centering
         \includegraphics[width=1\columnwidth]{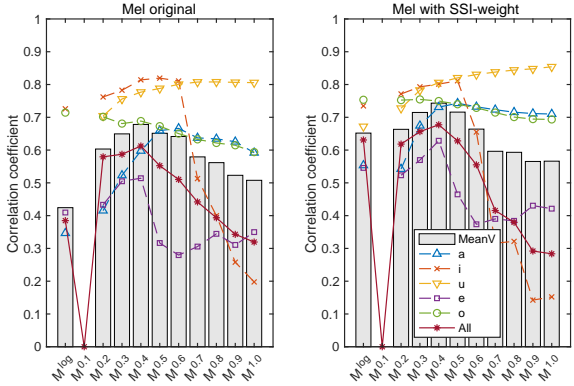}
                  \vspace{-15pt}
         \caption{Mel spectrum}
         \label{fig:CC-CompressMel}
     \end{subfigure}
           \begin{subfigure}[b]{\columnwidth}
         \centering
         \includegraphics[width=1\columnwidth]{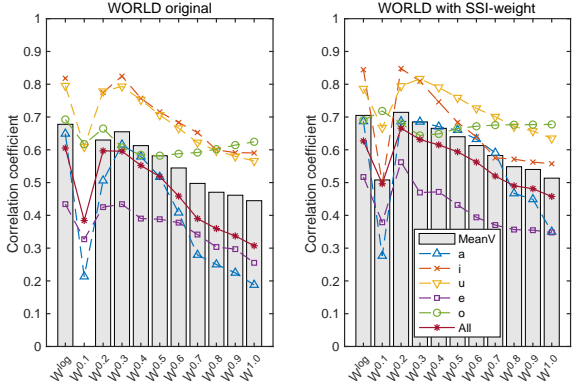}
                  \vspace{-15pt}
         \caption{WORLD spectrum}
         \label{fig:CC-CompressWORLD}
     \end{subfigure}
  \caption{ Correlation coefficients between the measured VTLs and the VTLs estimated from the log and power compressed spectrum. Left panel shows the result when using the original compressed spectrum and the right panel shows the result when using the spectrum with the SSI weight. Lines and bars are the same as in Fig.\,\ref{fig:EffectSSIhmax}.
  }
        \label{fig:CC-CompressSpec}
\vspace{-10pt}
\end{figure}

\subsection{Effect of compression on the estimation}
Initially, we sought to identify the log and exponential compressions that yielded the best VTL estimation in each spectrum. The resulting spectra are suitable for comparison with EP.

Figure\,\ref{fig:CC-CompressFourier} shows the correlation coefficients between the measured VTLs and the VTLs estimated from the various compressed Fourier spectra. The correlation coefficients were generally higher when using the spectrum with the SSI weight (left panel) than when using the original spectrum (right panel).The Fourier spectrum with the SSI weight $F_{SSI}^{log}$ yielded the best correlation coefficient and a relatively small variability between vowels. The $F^{log}$ in the left panel showed high correlation coefficient although it was not the best. We used these data for comparison.

Figure\,\ref{fig:CC-CompressMel} shows the results of the Mel spectrum.  The correlation coefficients were generally smaller than those in the Fourier spectrum. The Mel spectra of $M^{0.4}$ and $M_{SSI}^{0.4}$ were the best for each panel.
We also included $M^{log}$, which is the spectrum for the calculation of Mel-frequency cepstrum coefficient (MFCC)\,\cite{davis1980comparison}, and $M_{SSI}^{log}$ in the comparison.

Figure\,\ref{fig:CC-CompressWORLD} shows the results of the WORLD spectrum. We selected $W^{log}$ and $W_{SSI}^{log}$ for comparison, because they provided the best correlation coefficients on average.



\begin{figure}[t]
  \vspace{-10pt}
		\begin{center}
                \includegraphics[clip, width=0.9\columnwidth]
  {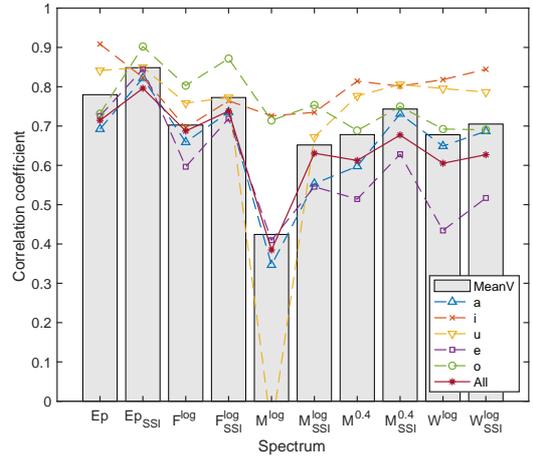}
             \vspace{-5pt}
               \end{center}
 	\caption{
  Correlation coefficients derived from different spectra. 
  Lines and bars are the same as in Fig.\,\ref{fig:EffectSSIhmax}
}        \label{fig:CorrTypeBarVowel}
  \vspace{-12pt}
\end{figure}


\subsection{Comparison between the EP and various spectra}

Figure ~\ref{fig:CorrTypeBarVowel} shows the correlation coefficients for the EP and the spectra selected above. 
$Ep_{SSI}$ had the highest correlation coefficient and 
smallest variability among the five vowels. In contrast, $M^{log}$ had the lowest correlation coefficient. 
Notably, the correlation coefficient was always higher in any spectrum when introducing the SSI weight.
Therefore, the SSI weight could improve the estimation of vocal tract information.

\begin{figure}[t]
		\begin{center}
                \includegraphics[clip, width=0.9\columnwidth]{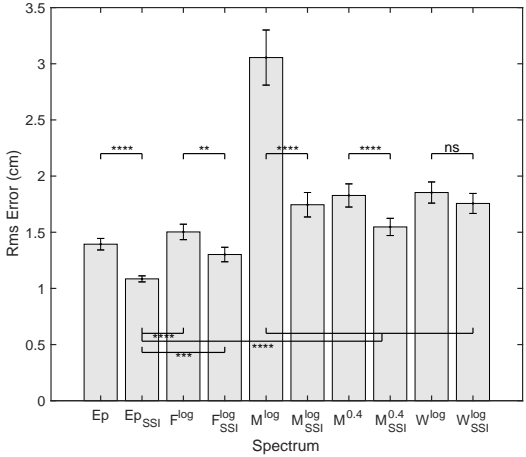}
                \end{center}
	\caption{Mean and standard deviation of the RMS errors in cm when the data of three speakers were randomly excluded 10 times. Tukey's HSD test: $*:p<0.05;  {**}:p<0.01; {***}:p<0.001; ****:p<0.0001$; ns: not significant. Note that there are many other significant differences, but they are not shown here for simplicity. }
\label{fig:RMSerrorStat}
  \vspace{-15pt}
\end{figure}

These differences were tested statistically. We performed the VTL estimation 10 times using data from 16 speakers after excluding two males and one female from the original 19 speakers at random. Subsequently, the RMS error between the measured and estimated VTLs was calculated. If the VTL estimation is sufficiently good, it should be stable and accurate even if three data points are randomly eliminated. Figure\,\ref{fig:RMSerrorStat} shows the mean and standard deviation and the results of Tukey's HSD multiple-comparison test. The RMS error for $Ep_{SSI}$ was approximately 1\,cm, which was significantly smaller than those for the other spectra. In contrast, the RMS error for $M^{log}$ was approximately 3\,cm, which was significantly greater than that for the other spectra.
More importantly, the results also demonstrate that introducing the SSI weight reduced the error in any spectrum. These reductions were statistically significant ($p<0.05$), except for the WORLD spectrum. It is noteworthy that the difference between the errors of $M^{log}$ and $M_{SSI}^{log}$ is extremely large. 





\subsection{Some lessons from the results}
Suggestions obtained from the results may serve as lessons for future speech signal processing.

\subsubsection{Effective auditory representation}

Many ``auditory motivated'' models have been proposed for various speech signal processing tasks. Most of them only introduced  peripheral frequency analysis, such as the auditory and Mel filterbanks. The current results imply that such a frequency analysis is not sufficient for extracting vocal tract information. The introduction of the SSI weight improved the performance. As shown in Figs.  \,\ref{fig:SSIweightMaking} and \ref{fig:SWMT_SWT_SSIweight}, the SSI weight is a simplified version of the Size-Shape Image in the SWMT, which was proposed as a computational theory of the central auditory process.
It is important to introduce knowledge of human auditory processing to derive effective auditory representation.

\subsubsection{Effectiveness of MFCC}
Mel-frequency cepstrum coefficient (MFCC) has been commonly used in  many kinds of speech processing after its tremendous success in ASR\,\cite{davis1980comparison}.
MFCC has also been used in VTL estimation 
\,\cite{lee1996speaker,pitz2005vocal,sanand2007linear,sarkar2009text}, even in recent DNN studies\,\cite{jaitly2013vocal,serizel2014vocal, tan2021vocal}.
However, it has rarely been questioned whether the MFCC is an effective representation for this purpose.

The log Mel-spectrum $M^{log}$ is a basic representation for calculating MFCC.  The results in Fig.\,\ref{fig:RMSerrorStat} imply that information about the vocal tract and glottal vibration is not sufficiently separated in $M^{log}$.  As the discrete cosine transform (DCT) is applied across all frequency ranges, all cepstrum coefficients unavoidably contain both types of information. Therefore, the use of MFCC does not seem effective for VTL estimation, even if a state-of-the-art DNN method is used in the back-end. This is because the DNN is required to segregate both types of information embedded in the individual cepstral coefficients before estimating the VTL. Although it would be possible to use a  large number of parameters to resolve this,  interpretation of the internal representation could be difficult because of complexity.
A modified version of the MFCC derived from the log Mel-spectrum associated with the SSI weight may improve the performance and interpretation.


\subsubsection{Merit and usage of the SSI weight}

the SSI weight is applicable to any type of commonly used spectra because it is a simple $F_o$ adaptive function on a frequency axis (Eq.\,\ref{eq:SSIweight} and Fig.\,\ref{fig:SSIweightMaking}).  It is not necessary to estimate $F_o$ accurately because the SSI weight is less sensitive to the $F_o$ value.  
the SSI weight can be easily implemented in any speech processing program by adding a few lines and an $F_o$ estimation package such as WORLD\,\cite{morise2016world}.
Practically, it has been introduced into an objective speech intelligibility measure, GESI\,\cite{irino2022speech,yamamoto2023intelligibility}, to improve the prediction of both male and female speech sounds.



\section{Summary}
In this study, we 
investigated auditory representations, which are effective for estimating vocal tract information. We proposed the use of the SSI weight which is derived from SWMT to segregate information about the vocal tract and glottal pulse from speech sounds.
The auditory EP associated with the SSI weight improved the estimation of VTLs measured from the MRI data.
Moreover, the estimation error was significantly smaller than when using the commonly used Fourier, Mel, and WORLD spectra.  It was also demonstrated that the SSI weight can be easily introduced into these spectra to improve the performance.


\section*{Acknowledgments}
This research was supported by JSPS KAKENHI  Nos. 21H03468 and 21K19794.
The authors would like to thank Prof. Kitamura for providing the female MRI-VTL data.


\begin{figure}[t]
  \begin{center} 
      \includegraphics[width = 1\columnwidth]{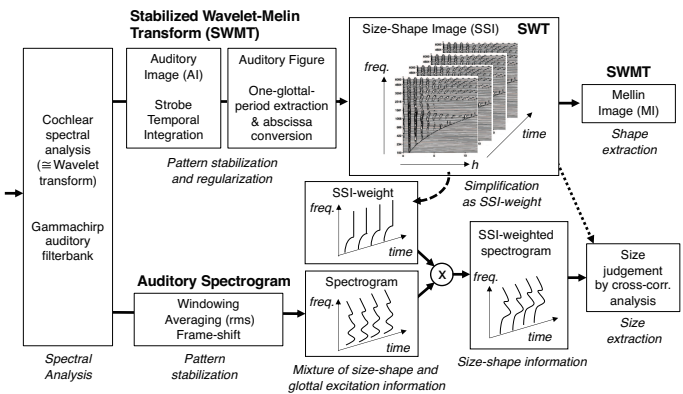}
	\caption{Relationship between SWMT, auditory spectrogram, and SSI weight\,\cite{matsui2022modelling}.
 The upper path shows the SWMT process; the lower path shows the auditory spectrogram process along with the SSI weight shown in the middle block. See Appendix A.}
	\label{fig:SWMT_SWT_SSIweight}
     \end{center} 
\end{figure}

\section*{Appendix A. SWMT and auditory spectrogram}
\label{sec:SWMT_SpectroAnalysis}


Figure ~\ref{fig:SWMT_SWT_SSIweight} shows the relationship between the SWMT (Stabilised Wavelet-Mellin Transform) \cite{irino2002segregating}, the auditory spectrogram, and the SSI (Size-Shape Image) weight introduced in Section \,\ref{sec:F0adaptiveWeightingFunc} (see Appendix A in \cite{matsui2022modelling} for more details).

The upper path in Fig.~\ref{fig:SWMT_SWT_SSIweight} shows the signal processing of the SWMT, which is supported by several experiments on size information processing in the auditory system (for example \cite{smith2005processing, ives2005discrimination, matsui2022modelling}).  This is effective in theory but has a problem when applied to practical applications.  The process of strobe temporal integration for stabilization and 2-dimensional conversion in SWMT is rather difficult to implement in a computational model. To resolve this problem, we developed the SSI weight (Eq. \ref{eq:SSIweight}) which is applicable to any spectrographic representation.

The lower path in Fig.~\ref{fig:SWMT_SWT_SSIweight} shows the analysis using an auditory spectrogram. The auditory filterbank is the same as that in SWMT. The output is processed by windowing and averaging to obtain an excitation pattern (EP). This process is simple and produces a stable representation of the spectral information in the sound, although it discards the temporal fine structure that also plays an important role in the auditory perception \cite{moore2008role}.
An auditory spectrogram is a stream of EPs derived from each frame. 
As described in Section\,\ref{sec:AuditoryProcess}, the product of the SSI weight (in the middle block) and the frame-based spectrum is suitable for extracting size information.






\printbibliography


\end{document}